\documentclass[fleqn,usenatbib]{mnras}

\usepackage{mathptmx}

\usepackage[T1]{fontenc}
\usepackage{ae,aecompl}

\usepackage{graphicx}	
\usepackage{amsmath}	
\usepackage{amssymb}	

\title[Compact jet in 4U 1543--47]{The appearance of a compact jet in the soft--intermediate state of 4U 1543--47}
\author[D. M. Russell et al.]{D. M. Russell$^{1}$\thanks{E-mail: dave.russell@nyu.edu}, P. Casella$^{2}$, E. Kalemci$^{3}$, A. Vahdat Motlagh$^{4,5}$, P. Saikia$^{1}$,
\newauthor
S. F. Pirbhoy$^{1}$, D. Maitra$^{6}$
\\
$^{1}$Center for Astro, Particle and Planetary Physics, New York University Abu Dhabi, PO Box 129188, Abu Dhabi, UAE\\
$^{2}$INAF, Osservatorio Astronomico di Roma, Via Frascati 33, I-00040, Monteporzio Catone (RM), Italy\\
$^{3}$Faculty of Engineering and Natural Sciences, Sabanc{\i} University, Orhanli-Tuzla, 34956, Istanbul, Turkey\\
$^{4}$Department of Physics and Astronomy, Texas Tech University, Box 41051, Lubbock, TX 79409-1051, USA\\
$^{5}$Istanbul Technical University, Faculty of Science and Letters, Physics Engineering Department, 34469, Istanbul, Turkey\\
$^{6}$Department of Physics and Astronomy, Wheaton College, Norton, MA 02766, USA \\
}

\date{Accepted XXX. Received YYY; in original form ZZZ}

\pubyear{2020}

\begin{document}
\label{firstpage}
\pagerange{\pageref{firstpage}--\pageref{lastpage}}
\maketitle

\begin{abstract}
Recent advancements in the understanding of jet--disc coupling in black hole candidate X-ray binaries (BHXBs) have provided close links between radio jet emission and X-ray spectral and variability behaviour. In `soft' X-ray states the jets are suppressed, but the current picture lacks an understanding of the X-ray features associated with the quenching or recovering of these jets. Here we show that a brief, $\sim 4$ day infrared (IR) brightening during a predominantly soft X-ray state of the BHXB 4U 1543--47 is contemporaneous with a strong X-ray Type B quasi-periodic oscillation (QPO), a slight spectral hardening and an increase in the rms variability, indicating an excursion to the soft--intermediate state (SIMS). This IR `flare' has a spectral index consistent with optically thin synchrotron emission and most likely originates from the steady, compact jet. This core jet emitting in the IR is usually only associated with the hard state, and its appearance during the SIMS places the `jet line' between the SIMS and the soft state in the hardness--intensity diagram for this source. IR emission is produced in a small region of the jets close to where they are launched ($\sim 0.1$ light-seconds), and the timescale of the IR flare in 4U 1543--47 is far too long to be caused by a single, discrete ejection. We also present a summary of the evolution of the jet and X-ray spectral/variability properties throughout the whole outburst, constraining the jet contribution to the X-ray flux during the decay.
\end{abstract}




\section{Introduction}

Matter accreting onto a black hole (BH) can, under some circumstances, result in collimated outflows that are accelerated to relativistic velocities. The circumstances -- the physical conditions required for these jets to be launched, is a hot topic of debate in accretion physics. For stellar-mass BHs accreting from a star in a binary system (BH X-ray binaries; BHXBs), recent efforts have led to global couplings between accretion inflow and outflow becoming apparent. More specifically, it was found that the radio and IR flux of jets is positively correlated with the X-ray flux \citep*{corbet00,corbet03,gallet03,homaet05,russet07,coriet09,gallet14,vincet18} in a fashion expected from theories of jet production in accreting BHs \citep*{falcbi95,heinsu03,kordet06}. However, it was also discovered that in some accretion state regimes \citep[which empirically reveal themselves as X-ray states defined by their spectral and variability properties; see][and references therein]{bell10} the jets are suppressed, possibly absent \citep[namely soft X-ray states; e.g.][]{fendet99,coriet11,russet11,Trusset19}. Prior to this quenching of the jets, bright radio flares are often reported, which are sometimes resolved into discrete ejecta \citep[e.g.][]{brocet02,millet12,rushet17}. Multi-band light curves of such flares can be modelled to solve for the energetics and velocities of the ejecta \citep[e.g.][]{tetaet17,fendmu16,fendbr19}.

\citeauthor*{fendet04} (2004, hereafter FBG04) attempted the first unified picture of jets in BHXBs, finding that most sources obey a specific pattern whereby the radio behaviour is linked to the luminosity and X-ray spectral state \citep[see also][]{miyaet95,maccco03}. The bright ejections and subsequent jet quenching are found to occur in approximately the same region of the X-ray hardness--intensity diagram (HID) for all known BHXB outbursts; namely the transition between the hard state and the soft state (FBG04). \citet*[][hereafter FHB09]{fendet09} extended this to timing behaviour, finding that the brightest discrete jet ejections occurred during the soft--intermediate state (SIMS), the low rms `zone' \citep[see also][]{bell10,mott16,fendmu16}, which is associated with quasi-periodic oscillations (QPOs).

As the X-ray spectrum softens over the transition, the timing properties change dramatically. Adopting the state classifications of \cite{bell10}, a BHXB typically enters the hard--intermediate state (HIMS) at the start of the transition, which exhibits a decrease in the fractional rms variability amplitude and a strong  QPO centred at $\sim 0.1$ -- 15 Hz \citep*[a type C QPO;][]{caseet05}. The transition to the HIMS is also associated with a fade of the IR emission from the jet \citep[e.g.][]{homaet05,coriet09,baglet18}, whereas the radio emission persists with a flat spectrum \citep[FHB09;][]{millet12,vandet13}. The source continues to soften and enters the SIMS, apparent from an abrupt drop in the fractional rms and either a type B QPO (at 5 -- 6 Hz) and/or a type A QPO (at $\sim$8 Hz).

It has been shown that the jet spectral break from flat/partially self-absorbed synchrotron to optically thin synchrotron, resides in the IR regime in the hard state \citep{corbfe02,gandet11,russet13a}. The break shifts to lower frequencies -- mm and radio -- during the transition through the HIMS, to the SIMS, and shifts back to the IR on return to the hard state \citep{vandet13,corbet13,kaleet13,russet13b,Trusset14,baglet18}, and there is a correlation between the X-ray hardness and the jet break frequency \citep{koljet15}.
The radio flares, which are optically thin, are observed over the transition to the SIMS, during the SIMS, over the transition from the SIMS to the soft state, and/or during transition from the soft state to the `ultrasoft' state \citep[e.g. FHB09;][]{kaleet16,Trusset19}.

The exact time the core radio jet is quenched is however unclear, because the bright radio emission could result from collisions between discrete ejecta and the environment or the slower, hard state jet \citep[see discussion in][]{Trusset19}. If the core jet itself could be observed to quench or recover, with simultaneous X-ray monitoring this would recover the exact position of the `jet line' in the HID. The changes in X-ray properties associated with this would shed light on the physical process that prohibits jet production in accreting black holes.

Here we report the discovery of a compact jet in the SIMS, contemporaneous with a type B QPO, in the BHXB 4U 1543--47 during its 2002 outburst. The result is derived from multiwavelength observations; the outburst was monitored extensively at X-ray, optical/IR (OIR) and radio wavelengths.

\begin{figure}
\centering
\includegraphics[width=\columnwidth,angle=0]{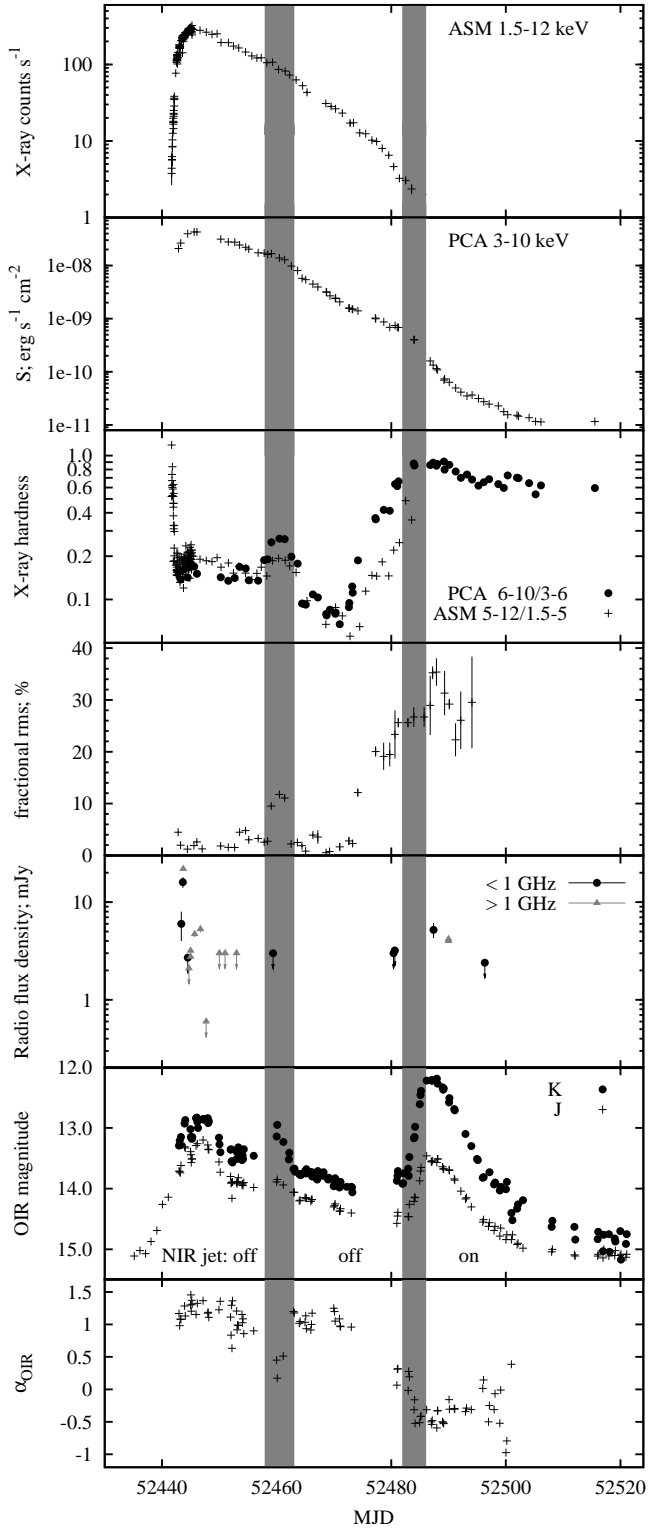}\\
\caption{Light curves of the 2002 outburst of 4U 1543--47.  The shaded regions in the light curves indicate when the OIR jet is quenching/recovering. The X-ray light curves, hardness and fractional rms variability are shown in the upper four panels; the radio flux density, IR magnitudes, and de-reddened $J$ to $K$ spectral index are shown in the lower three panels. The jet IR emission is quenching or recovering during the times of the grey shaded regions (see text).}
\label{lc1}
\end{figure}

4U 1543--47 (IL Lup) is a transient BHXB \citep{oroset98}, which has performed four outbursts since its discovery in 1971 \citep{matiet72,kitaet84,buxtba04}. The distance to 4U 1543--47 is $7.5 \pm 0.5$ kpc \citep{oroset02}. The orbital period is determined to be $26.79377 \pm 0.00007$ hrs, and the inclination angle is estimated as $20.7^{\circ} \pm 1.5^{\circ}$ \citep{oros03}. \cite{russet06} compiled from the literature a central BH mass of $9.4 \pm 1.0 ~M_{\odot}$ and a companion mass of $2.45 \pm 0.15 ~M_{\odot}$ for this system \citep[see also][]{saiket19}. Its 2002 outburst was monitored regularly with the X-ray satellite  Rossi X-ray Timing Explorer \citep*[RXTE;][]{parket04,kaleet05,kaleet13,reiget06,gierne06,glioet10,koen13,mornmi14,dincet14,lipuma17,vahdet19} and with ground based optical and IR \citep{buxtba04} and radio \citep{parket04,kaleet05} facilities. As such this outburst comprises one of the richest datasets of the multiwavelength evolution of a transient BHXB. The synchrotron emission from the jets produced in this BHXB accounts for the radio emission, and was also found to contribute to (and sometimes dominate) the OIR regime \citep{buxtba04,kaleet05,russet07,russet13a}.

\begin{figure*}
\centering
\includegraphics[width=12cm,angle=270]{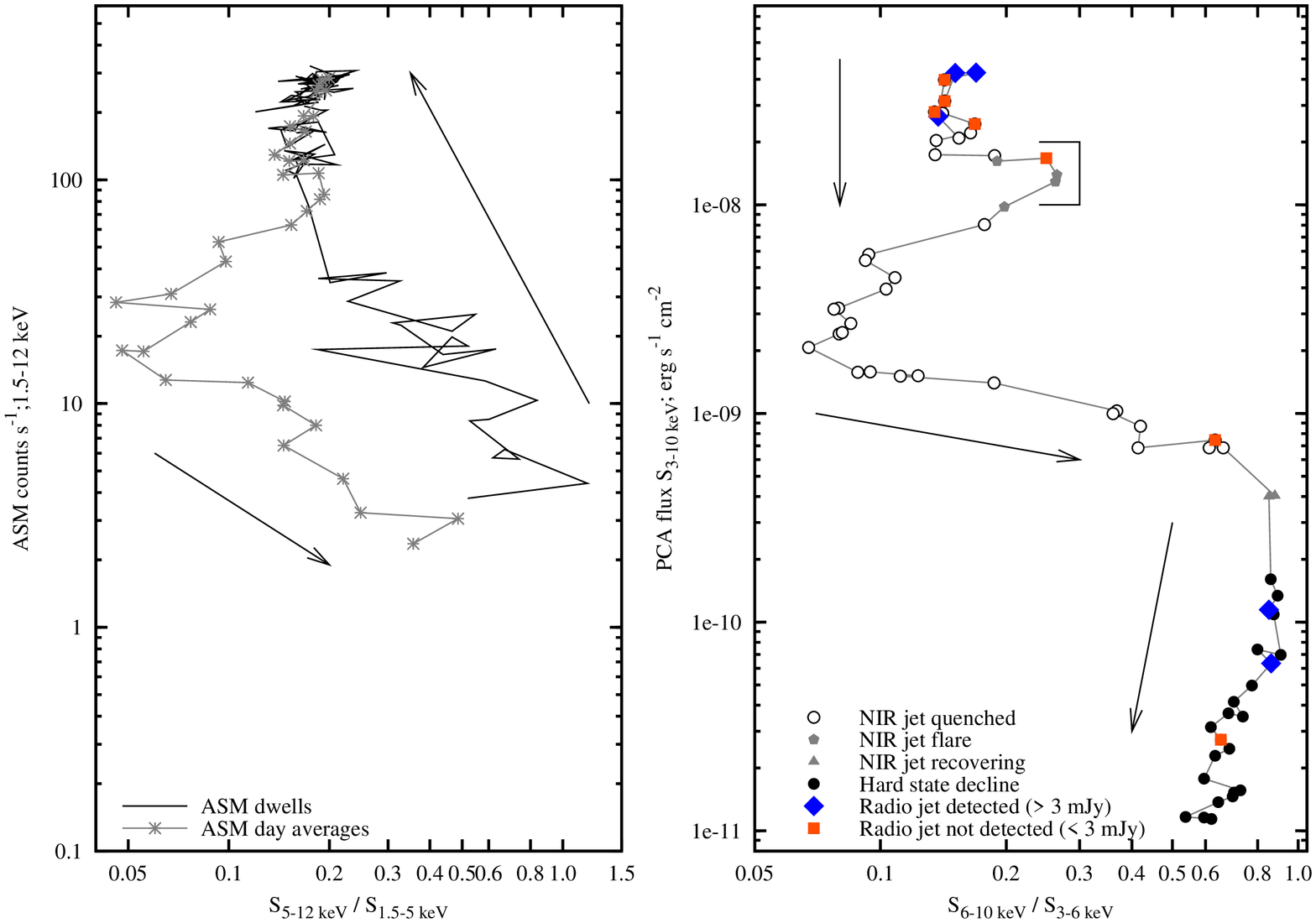}\\
\caption{X-ray hardness--intensity diagrams (HIDs) of the outburst of 4U 1543--47. \emph{Left:} HID constructed from RXTE ASM data, showing the initial hard state rise. \emph{Right:} HID constructed from RXTE PCA data, with symbols indicating the jet properties from IR and radio data. The open box near the top centre illustrates the brief excursion to the SIMS. The arrows show the direction of evolution of the outburst.}
\label{hid}
\end{figure*}

\section{Data collection}

All the multiwavelength data for this outburst exist in the literature. Radio observations are taken from \cite{parket04} and \cite{kaleet05}. The source was observed at 843 MHz with the Molonglo
Observatory Synthesis Telescope (MOST), at 617, 1027 and 1287 MHz with the Giant Metrewave Radio Telescope (GMRT), and at 4.80 and 8.64 GHz with the Australia Telescope Compact Array (ATCA). The source was detected at radio frequencies during the first week of the outburst and during the hard state decay in the latter stages of the outburst. Upper limits were acquired at other times, although many were not very constraining, with typical upper limits being $<$ a few mJy \citep[for details see][]{parket04,kaleet05}.

$J$-band ($1.2 \mu$m) and $K$-band ($2.2 \mu$m) NIR magnitudes are taken from \cite{buxtba04}. In their paper, \citeauthor{buxtba04} present monitoring in three optical ($B$, $V$, $I$) and two NIR ($J$, $K$) filters. Since we are interested in the jet component of the emission, which is much more prominent in the NIR compared to the optical, we use the $J$ and $K$-band magnitudes in the following analysis (these two filters also had the most complete coverage of the outburst).

Since 4U 1543--47 contains a bright star of magnitude $J \sim K \sim 15$ mag \citep{buxtba04}, the flux from this star makes a considerable contribution to the NIR magnitudes during the outburst, especially at low flux levels at the start and end. To derive the NIR emission from the accretion flow (disc and jet) in these wavebands ($J_o$ and $K_o$) we therefore subtracted the known flux of the star from these data, which was assumed to be constant. Finally, the intrinsic fluxes and NIR spectral indices were calculated by de-reddening the magnitudes adopting an interstellar extinction of $A_{\rm v} = 1.55$ \citep[which is fairly well determined;][]{oroset98} and the extinction law of \cite*{cardet89}.

The RXTE Proportional Counter Array (PCA) fluxes and X-ray photon index values are taken from \cite{dunnet10}. At the start of the outburst, the All-Sky Monitor (ASM) on board RXTE monitored the rapid rise of the flux before the PCA was triggered. The first PCA observations showed 4U 1543--47 was already in transition towards the soft state \citep[e.g.][]{reiget06}, whereas the hard state rise was seen with the ASM \citep{gierne06}. We therefore include publicly available ASM data. Both PCA and ASM monitored the progress of the outburst for almost all of the outburst, with few gaps in the coverage. RXTE PCA X-ray variability properties are taken from \cite{kaleet13}; see also \cite{dincet14}. The data included in those papers were from the latter stages of the outburst; here we extended the analysis to the whole outburst \citep[for details of the X-ray timing analysis, see section 2.2 of][]{kaleet13}. QPO classification was taken from \citet{parket04} and from \citet{gaoet17}, and further refined for the observations performed during the SIMS.

\begin{figure}
\centering
\includegraphics[width=\columnwidth,angle=0]{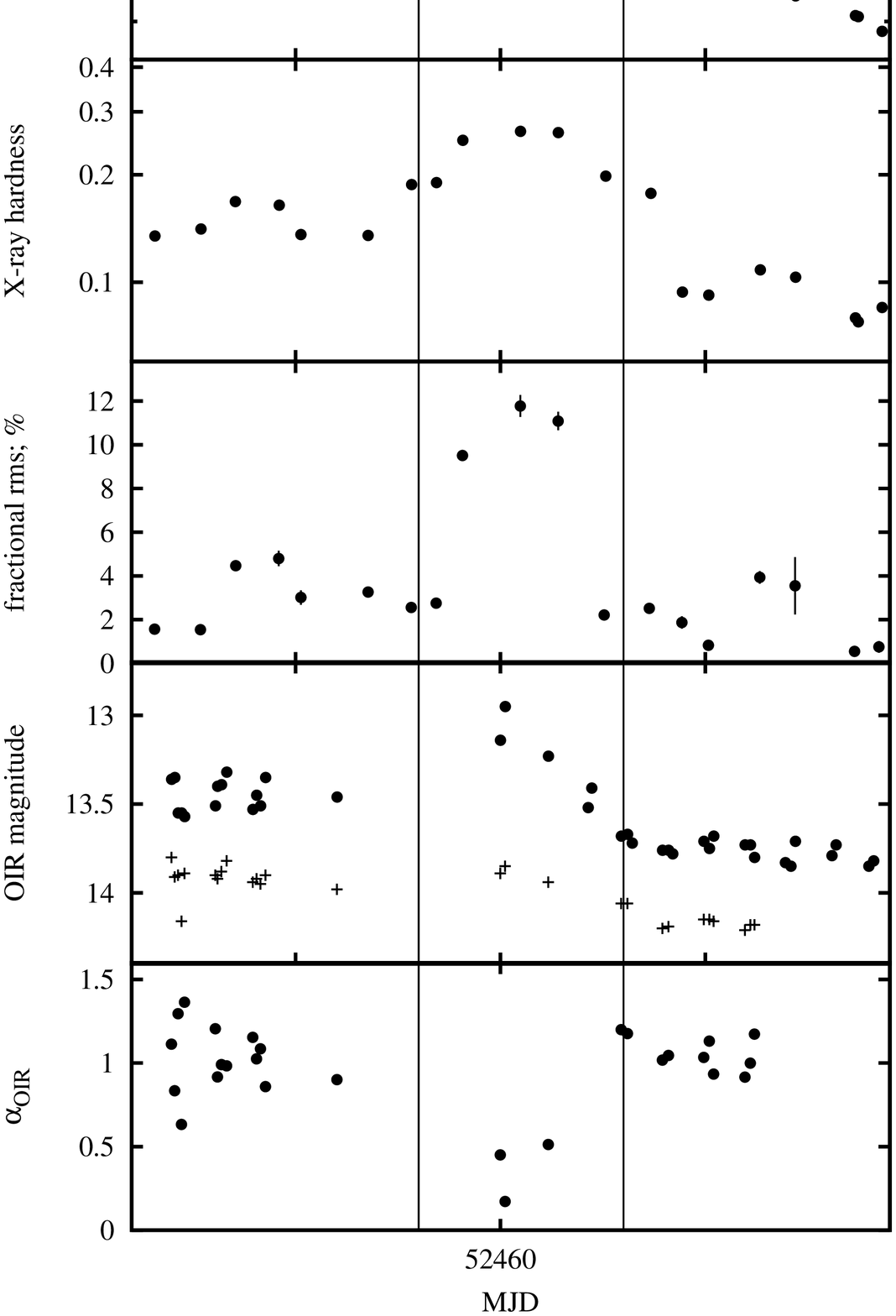}\\
\caption{Light curves, similar to Fig. \ref{lc1}, focusing on the time of the IR flare in the SIMS.  The vertical dotted lines encompass the flare, and are the same dates as the start and end of the left shaded region in Figs. \ref{lc1} and \ref{lcjet}. The RXTE PCA X-ray light curve, hardness and fractional rms variability are shown in the upper three panels; the IR magnitudes and de-reddened $J$ to $K$ spectral index are shown in the lower two panels.}
\label{lczoom}
\end{figure}

\section{Results and analysis}

\subsection{Multiwavelength evolution of the outburst}

We present the light curves of the whole outburst at all wavelengths collected for this study in Fig. \ref{lc1}. The X-ray flux (or count rate) and hardness from RXTE ASM and PCA are plotted in the upper three panels. ASM daily averages are shown after MJD 52445.3. Before this date, ASM individual observations are used due to the rapidity of the outburst rise and the high signal-to-noise ratio (S/N; the source peaked at $\sim 320$ ASM counts s$^{-1}$; 1.5--12 keV). For PCA the hardness is defined as the 6--10 keV flux divided by the 3--6 keV flux. For ASM it is the 5--12 keV count rate divided by the 1.5--5 keV count rate. The X-ray fractional rms variability light curve is plotted in the fourth panel. The radio flux, OIR observed magnitude and NIR intrinsic (de-reddened) spectral index are presented in the lower three panels. The grey shaded regions mark the epochs (MJD 52458--52463 and 52482--52486) in which there is evidence for a quenching or recovering of the jet emission observed in the OIR regime (see below).

It is clear that the source began its outburst with a hard X-ray spectrum, as was seen from the RXTE ASM hardness ratio. The X-ray flux rapidly peaked and softened, then subsequently began a slow decay, with the X-ray spectrum remaining soft until gradually hardening half way through the outburst \citep[for detailed spectral analysis see e.g.][]{parket04,kaleet05,reiget06,dunnet10,glioet10,koen13,lipuma17}. In Fig. \ref{hid} the X-ray HIDs of the outburst from RXTE ASM (left panel) and PCA data (right panel) indicate that the source followed the usual anti-clockwise loop in the HID, common to BHXBs (e.g. FBG04; FHB09). The X-ray variability was weak (rms $\lesssim 6$ per cent) in the soft state, increasing to $\sim 10$--35 per cent in the hard state decay.

During the transitional states at high luminosity several radio flares were reported, which are also shown in Fig. \ref{lc1}. The epochs of radio detections and non-detections are indicated in the HID in Fig. \ref{hid} (right panel). The initial radio flares were seen at a time when the X-ray spectrum was already quite soft \citep[MJD 52443.4, when the X-ray photon index was $\Gamma = 2.5$--2.7;][]{parket04,dunnet10}. During the hard-to-soft transition (near the outburst flux peak), the `jet line', when bright discrete ejecta are launched and the core jet is quenched, is usually at or around the SIMS--soft state transition (FHB09). Here, the hard-to-soft transition in 4U 1543--47 was very rapid, and the first radio observations were taken when the source was already in the soft state. The bright radio flares at this time may originate either in the core jet or in shocks downstream in the flow (internal shocks or interactions with the interstellar medium).

During the transition back to the hard state at lower luminosity, the location of the jet line is debated (FHB09). In 4U 1543--47, a radio non-detection exists just before the source joined the canonical hard state in the HID, and radio detections were made once 4U 1543--47 was fully back in the hard state. This implies that in this source, the `jet line' during the soft-to-hard transition lies at a fairly hard X-ray hardness, in fact consistent with when the source was fully back in the hard state (FHB09). However, we note that the radio upper limits of $< 3$ mJy before this are not particularly constraining. For 4U 1543--47 this radio brightening occured at an X-ray hardness ratio about four times harder than the jet line during the hard-to-soft transition (Fig. \ref{hid}, right panel). In GX 339--4, which had good radio coverage of this transition during its 2010--2011 outburst, the radio jet also switched on when the source entered the hard state -- similar to 4U 1543--47. In GX 339--4 this occured just a few days before the start of the IR brightening \citep{corbet13}. On the other hand, the jet line may appear to be at a softer hardness than this in some sources such as XTE J1720--318 (see discussion in section 7 of FHB09). Sensitive radio monitoring during the soft-to-hard transition will be needed to assess when and how the jet returns during this transition in compilations of sources \citep[see also][for a discussion]{kaleet13}.

The OIR flux evolution is discussed in detail in \cite{buxtba04}. In the lowest panel of Fig. \ref{lc1} the intrinsic (de-reddened) NIR ($J$-band to $K$-band) spectral index is presented. Throughout the first half of the outburst, the NIR colour is generally blue, with a spectral index typical of the outer regions of an accretion disc \citep[which is likely irradiated;][]{hyne05,russet07}; $\alpha \approx 1$ (where $F_{\nu}\propto \nu^{\alpha}$). Once the X-ray spectrum has hardened in the second half of the outburst, a dramatic brightening is seen in the NIR light curves ($\sim 2$ mag in $K$-band) and less so in the optical light curves. The origin of this IR flare was explained \citep{buxtba04} to be different from the primary outburst, very likely synchrotron emission from the jet in 4U 1543--47 \citep[see also][]{russet07,russet13a,kaleet13,dincet14}. Using the optical amplitude, this brightening is classed as a reflare according to the classification of \cite{zhanet19}. The NIR spectral index migrates from $\alpha \approx +1$ in the soft state to $\alpha \approx 0$ to $-0.5$ in the hard state outburst decay.

\subsection{A brief and unexpected IR flare in the SIMS}

There is a brightening of the NIR flux (and reddening of the spectral index) around MJD 52460, lasting a few days. This flare is much brighter at IR, compared to optical, wavelengths \citep{buxtba04}. The flare is accompanied by a brief hardening of the X-ray flux and increase of the rms to $\sim 10$ per cent (Figs. \ref{lc1} and \ref{lczoom}). Before and after this NIR brightening, the source was in a soft state. The hardening, increased X-ray variability, and the appearance of a prominent type B QPO \citep[see fig. 8(b) in][]{parket04} at this time indicates that the source was in the SIMS \citep[or the `steep power law state', in the definitions of][]{mcclre06}. The left shaded region in Fig. \ref{lc1} encompasses the date range of the flare (MJD 52458--52463), limiting the duration of the flare to $< 5$ days. In Fig. \ref{lczoom} we show this region of the light curves in detail. A weak type A QPO is first detected on MJD 52457 at a frequency of $\nu_{\rm QPO} \sim 10.1$ Hz, observed again with similar properties on the following day. A type-B QPO is then observed during the three observations performed on the MJD interval 52459--52461 - corresponding to the local peak in X-ray hardness, with a centroid frequency $\nu_{\rm QPO} = 7$--8 Hz and weak harmonic features at $0.5 \nu_{\rm QPO}$ and $2 \nu_{\rm QPO}$, before disappearing on MJD 52462 \citep{parket04}. The excursion to the SIMS was brief, lasting only $\sim 5$ days -- contemporaneous with the IR flare.

The location of the SIMS and IR flare in the HID is indicated by the open box region in Fig. \ref{hid}, right panel. The IR flare flux peaked on MJD 52460, at a de-reddened $K$-band flux density of $F_{\nu} = 5.2$ mJy, of which the flare itself contributes 2 mJy. There is a single radio observation during the SIMS, on MJD 52459, yielding a radio upper limit of $< 3$ mJy at 843 MHz. If it is the compact jet we would expect a radio flux $\leq \sim 2$ mJy if the spectrum is flat or slightly inverted, so the radio upper limit is consistent with a compact jet. It is not surprising that the radio emission was not detected as the compact jet could have a slightly inverted spectrum from radio to IR; indeed, during the hard state decay the radio spectrum was inverted, with $\alpha = +0.08$ \citep{kaleet05}.

The IR excess emission (above the disc component in the spectral energy distribution), which is commonly attributed to synchrotron emission from the compact jet, is usually associated with the hard state \citep[e.g.][]{miglet07,buxtet12,kaleet13,russet06,russet13a}. The IR excess usually fades during the hard to HIMS transition \citep[e.g.][]{coriet09,cadoet11,saiket19}, with additional brief IR flares only seen close to the hard--HIMS transition if the source does not make a smooth transition \citep{baglet18}. During the SIMS and soft state the IR emission is quenched as the jet spectrum evolves \citep[e.g.][]{russet13b,corbet13}. The only exception is the observation of brief IR jet flares from discrete ejections in sources that do not perform the usual path through the HID, such as GRS 1915+105 and V404 Cyg \citep{fendet97,shahet16}. The IR flare of 4U 1543--47 during the SIMS is therefore the first compact jet detected at IR wavelengths in the SIMS (as far as we are aware).

\begin{figure}
\centering
\includegraphics[width=\columnwidth,angle=0]{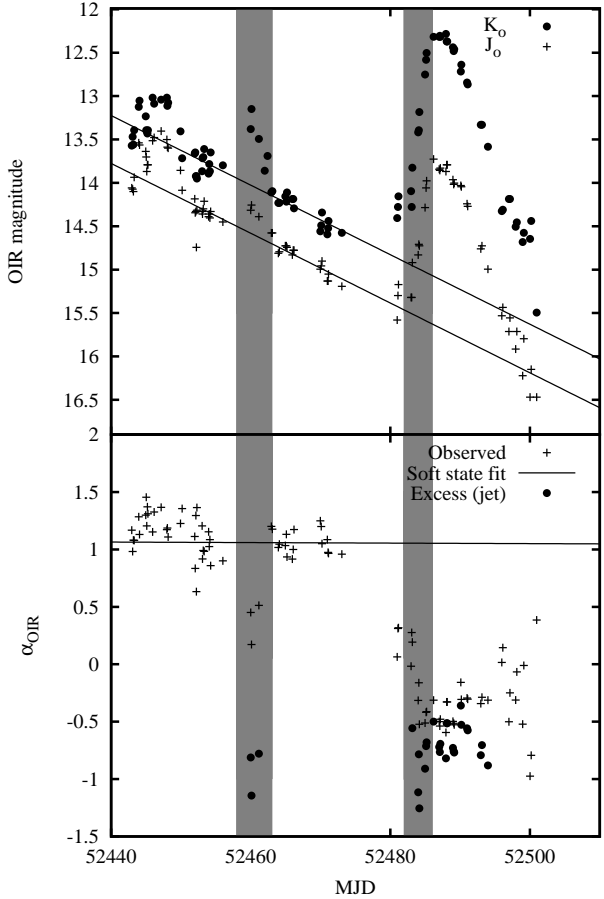}
\caption{\emph{Upper panel}: De-reddened IR light curves of 4U 1543--47 during the 2002 outburst, showing exponential decay fits to the soft state data (black solid lines). \emph{Lower panel}: Intrinsic NIR spectral index of the different emitting components. In the soft state (with no jet emission) the mean NIR spectral index is $\alpha = 1.1$. The excess emission above this disc component has a spectral index of $\alpha = $ -1.2 to -0.8 (brief flare in the SIMS) and $\alpha = $ -1.3 to -0.3 (during the hard state decay). The grey shaded regions are the same as in Fig. \ref{lc1}.}
\label{lcjet}
\end{figure}

\subsection{Separating the NIR disc and jet emission}

The NIR light curves are again plotted in Fig. \ref{lcjet} (upper panel) after subtraction of the light from the companion star. During the soft state, the NIR light curves appear to be approximated by an exponential decay. Following the method of \cite{russet10}, we fit this decay in each of the two bands and estimate the excess above this decay in the hard state after the transition by subtracting the light from the extrapolation of this decay \citep[see also][]{dincet12,russet12}. We define the soft state as between MJD 52452 and 52473 and fit these data, excluding the data at MJD 52458--52463 when the source made the aforementioned brief excursion to the SIMS \citep[see also fig. 2 in][]{russet13a}. We find a soft state fit decay rate of $40 \pm 2$ milli-mag d$^{-1}$ in $J$-band and $40 \pm 3$ milli-mag d$^{-1}$ in $K$-band (shown as solid black lines in Fig \ref{lcjet}, upper panel).

The SIMS stands as a clear $\sim 4$ day NIR excess above the decay in the soft state, most prominent in $K$-band. In the soft state, no jet emission is expected, and indeed the spectral index is typical of the outer accretion disc ($\alpha \sim 1$). Later, the reflare in the hard state decay is $\sim 2$ mag and $\sim 3$ mag above the extrapolation of the soft state exponential decay in $J$ and $K$-bands, respectively. Near the end of the outburst the NIR fluxes fade to a level close to the extrapolation of the soft state fit. Under the assumption that the emission from the disc continued to decay following the fit to the soft state (which is expected for fast-rise, exponential decay outbursts like this one), we measure the flux of the excess above this decay by subtracting the extrapolation of the fit in each band. We cannot rule out the possibility that the disc fade deviated from this exponential decay, but the data at the end of the outburst lie close to the extrapolation, which supports this assumption. We do not plot the data after MJD $\sim 52500$ because after this date the observed IR flux reaches close to the predicted disc flux extrapolation from the hard state, and it also approaches the quiescent level, so the method of measuring the jet component in this way becomes unreliable, with large errors after this date. The exponential decay fits in Fig. \ref{lcjet} appear to differ slightly from those of fig. 2 in \citet{russet13a} but in that figure the apparent magnitudes ($J$, $K$) are shown whereas here we plot the de-reddened, companion star-subtracted magnitudes ($J_o$, $K_o$) in Fig. \ref{lcjet}. Since the amplitude of the IR excess is so large during the decay, the disc contribution is small enough that deviations from this extrapolation will likely have small effects on the estimates of the jet spectral index, except possibly at late times, when the jet flux fades close to the disc extrapolation (MJD $\sim$ 52500).

From the flux of the excess above the exponential decay, the spectral index of the excess emission is estimated, and shown in the lower panel of Fig. \ref{lcjet} (black solid circles). In the hard state outburst decline the spectral index of the excess is $\alpha = -0.7 \pm 0.3$ \citep[see also figs. 1(e) and 2 of][]{russet13a}. This is typical of optically thin synchrotron emission from the steady, compact jet. Some of the first few observations of the excess have a slightly redder spectral index; $\alpha \sim -1.2$. This is similar to what was seen in the BHXB XTE J1550--564 -- in that source the spectral index increased from $\alpha \sim -1.5$ to $\alpha \sim -0.6$ in 20 d, which may result from a gradual increase in the particle energies in the flow \citep{russet10}. In 4U 1543--47 we here see a much more rapid change in spectral index, converging to $\alpha \sim -0.7$ in 1--3 d.

Near the start of the outburst, no significant excess is seen, indicating a probably negligible jet contribution. However, we note that the first $K$-band data in the light curve were after the rapid X-ray softening seen by RXTE ASM. The IR excess during the inital hard state rise of the outburst may have therefore been missed. The $\sim 4$ day NIR flare around MJD 52460 is however much more apparent. Five $K$-band data points and three $J$-band lie clearly above the exponential decay. By subtracting the exponential decay of the disc component we measure (Fig. \ref{lcjet}, lower panel) the spectral index of the NIR flare from these data to be $\alpha = -0.68 \pm 0.23$. This is also typical of optically thin synchrotron emission, and strongly suggests a brief, four day reappearance of the compact jet in the SIMS. The extrapolation of the 2 mJy IR optically thin synchrotron spectrum to radio frequencies severely overpredicts the observed radio upper limit, which implies the spectral break in the jet spectrum must reside somewhere in the mm to IR part of the spectrum.

\section{Discussion}

\subsection{Origin of the infrared flare in the SIMS}

The spectral index, and light curve morphology, of the IR flare give clues to its origin. Above we establish that since its spectral index is $\alpha = -0.68 \pm 0.23$, it is likely due to optically thin synchrotron emission. A hot inflow close to the black hole could produce OIR synchrotron emission \citep*{veleet13}, but the spectrum of such a component is expected to be flat ($\alpha \sim 0$) at optical wavelengths, getting fainter in the IR ($\alpha \gg 0$). The optically thin part of the synchrotron spectrum of the hot flow ($\alpha < 0$) is expected at UV wavelengths. We can therefore rule out this component as the origin of the IR flare in the SIMS, and also the IR excess during the hard state decay, for the same reason. Since the dust extinction towards 4U 1543--47 is both low and well constrained, uncertainties in the IR spectral index from extinction are negligible, and cannot bias our spectral index estimates for this source.

Several lines of reasoning support the origin of the IR flare to be the core, compact jet, rather than discrete ejecta or collisions downstream in the jet. At radio frequencies, a single optically thin synchrotron flare typically rises quickly, peaking at several mJy, then fades on timescales of hours to days--weeks. This has some similarities to the IR flare seen here -- similar peak flux density and flare timescale. However, radio flares that are optically thin should be much fainter at IR frequencies, by $\sim 3$--4 orders of magnitude in flux density for $\alpha = -0.7$. Consequently, a radio flare peaking at 20 mJy (e.g. the brightest seen from 4U 1543--47; Fig. \ref{lc1}) would be $\sim 5 \mu$Jy at IR $K$-band. It is not then surprising that to date, no optically thin radio flares seen over state transitions have been reported to have an IR counterpart. The only exceptions are the OIR and radio flares seen in GRS 1915+105 and V404 Cyg \citep{fendet97,shahet16}, but these show a delay between bands (i.e. it is not from one single power law spectrum) and occur on minute--hour timescales, and have similar flux densities in IR and radio.

By comparison with XTE J1550--564, the resolved X-ray and radio ejecta seen on arcsec--arcmin scales by \cite{corbet02,miglet17} in that source were not detected in IR, and by interpolating between radio and X-ray, \cite{corbet02} showed that the IR would be fainter than is detectable by the Very Large Telescope, i.e. several orders of magnitude fainter than our bright IR flare for 4U 1543--47. The radio to X-ray spectra of the resolved, extended jets of XTE J1550--564 were consistent with a single power law, except on some dates when a broken power law was favoured, with a spectral break at $\sim 10^{15}$ Hz, a slightly higher frequency than IR. In addition, if the observed IR flare was due to shocks in the jet downstream or interactions between the jet and the interstellar medium, we would expect the optically thin IR spectrum to extend to radio frequencies. In that case we would expect the flare (2 mJy in IR $K$-band) to peak at $\sim 8$ Jy at 1 GHz (assuming $\alpha = -0.7$). This is clearly ruled out by the radio observations, with a radio upper limit of $< 3$ mJy at the time of the IR flare. To summarize, the IR flare and its inferred synchrotron spectrum is far brighter than any extended, resolved ejecta seen before from other sources.

The combination of the NIR spectral index and the radio upper limit implies a broken power law for the synchrotron spectrum, i.e. a compact jet. In the compact jet scenario, the IR emission -- if optically thin -- is produced by the accelerated particles in the collimated outflow that are closest to the jet base, at distances $\sim 10^2$--10$^4$ gravitational radii ($r_{\rm g} = G M / c^2$) from the black hole \citep*[e.g.][]{market01,market03,ceccet18}. This size scale has been confirmed from rapid variability timescales and IR--X-ray cross-correlations, and corresponds to $\sim 0.1$ light-seconds \citep[e.g.][]{caseet10,kalaet16,gandet17,malzet18,vincet19,paicet19}. Since the size scale of the IR emitting region is on the order of 0.1 light-seconds, and the flare lasted several days, this cannot be due to a single discrete plasma ejection, as that would last for $< 1$ sec at IR frequencies. In addition, the IR flare is simultaneous with the X-ray hardening and 6 Hz QPO that has been thought to be associated with the `jet line', so it is plausible that the compact jet switched on when the source crossed the jet line from the soft state to the SIMS.

Above, we find that the data imply that the jet spectral break must reside somewhere in the mm to IR part of the spectrum during the IR flare. During the SIMS, the X-ray photon index is $\Gamma = 2.5 \pm 0.2$ \citep{parket04,dunnet10}. From the correlation between photon index and jet spectral break frequency \citep{koljet15}, this would correspond to a jet break between $10^{9.4}$ and $10^{11.0}$ Hz, i.e. in the mm regime. It is also possible that this brief compact jet `flare' does not obey the photon index -- jet break correlation for some unknown reason.

\subsection{Association of the compact jet with the SIMS and the Type B QPO}

The more common type C QPO (not seen in the SIMS) has been associated with Lense-Thirring precession in the inner accretion flow \citep{stelvi98,ingret16}, a process which could rapidly change the jet orientation if it is launched from a precessing flow \citep{millet19}.

A possible association between the X-ray type-B QPO and the so-called jet line as defined with radio observations, has been proposed by several authors in the literature \citep*[e.g.][]{soleet08}, although FHB09 showed that this association is all but straightforward. The main obstacle in studying this association is the sparse nature of most radio data, and more intrinsically the time delay between any transition or event happening in the inflow and the possibly related behaviour of the radio emission, coming from far along the jet. Our IR flare allows a precise location of the jet line between the SIMS and the soft state; this has been unattainable from radio observations because discrete ejections radiating at radio frequencies could have been launched several days previously and brightened due to internal shocks or collisions with interstellar gas.

In only a few cases has it been possible to pinpoint the actual epoch of the discrete ejection, by tracking the motion of the resolved radio blobs \citep{millet12,rushet17,Trusset19} or by multi-frequency modelling of time-resolved flares \citep{tetaet17}. The ejection epoch appears to correspond to just {\it before} the appearance of the type-B QPO, suggesting that what triggers the discrete ejecta is perhaps the drop in X-ray rms \citep[see also][]{radhet16}, more than a specific feature in the X-ray power spectrum.

At the same time, there is now growing evidence for the type-B QPO to originate in a jet-like structure, as suggested by several authors based on independent arguments \citep*{mottaet15,sriramet16,stevutt16,deruiteret19}. The clear association we report here between a compact jet and the type-B QPO, strongly supports this, while at the same time opening new questions. The first such question is whether this result is representative of all sources or not. This is one of the best IR coverages of the hard-to-soft spectral transition of a BH transient, and especially of the unusual behaviour of crossing of the jet line into the SIMS from the soft state (i.e. a brief hardening). Future, dense IR monitoring of this phase of other BHXB outbursts will allow us to answer this question.

Assuming both our result and those by \citet{millet12}, \citet{rushet17} and \citet{Trusset19} are representative of all sources, the temporal sequence is then as follows: (1) the IR flux from the steady compact jet is seen to quench rapidly at the transition from the hard state to the HIMS, eventually disappearing during the HIMS; (2) at the end of the HIMS when the X-ray rms drops, or during the SIMS or SIMS-to-soft transition, discrete ejections happen and cause the observed (delayed) radio flares; (3) the compact jet is quenched in the soft state, but may be present in the SIMS. In 4U 1543--47, the compact jet appears when a transition back to the SIMS occurs, from the soft state. A question remains open then on what is the connection between the discrete ejections and the compact jet observed soon after. To answer this question we will need again further data, in particular dense multi-wavelength monitoring in X-rays, IR and radio, to track all the components (i.e. X-ray variability, IR emission from a compact jet, and radio emission from discrete ejections) and study the details of their temporal association.

It is possible that the reason an IR flare has not been seen in other BHXBs at this stage of the outburst may be because most BHXBs do not make a transition from the soft state to the SIMS at high luminosity \citep[one other example is MAXI J1659--152, but no radio or IR data were available during those transitions;][]{vandet13}. Most BHXBs transition in the opposite direction -- SIMS to soft state -- and the jet quenches. Our result clearly supports the association between the SIMS and the activation of the jet. It is unclear however, why an IR jet is not detected in the SIMS in the opposite transition (HIMS to SIMS to soft state). Perhaps the discrete ejections, launched just prior to this SIMS, somehow prevent the compact jet to return or to radiate at IR wavelengths, during this transition.

At lower luminosities, BHXBs make a transition from the soft state to the SIMS, on their way to the hard state, during outburst decay. No IR flares have been reported during this stage of the outburst from the soft state to the SIMS. Type B QPOs have been reported in the SIMS on the way back to the hard state, but only in GX 339--4 \citep{mottet11}, not in other sources \citep[e.g.][]{vahdet19}. Generally the IR flux of the compact jet rises after the radio, as the spectrum evolves during the transition to the hard state \citep{kaleet13,corbet13,russet13b}. More observations of this transition would help to test this further, if a QPO is present during this transition, and if an IR flare occurs at all, before the compact jet builds up during the HIMS and hard state decay.

\subsection{Jet contribution to the X-ray luminosity}

By disentangling the IR disc and jet contributions, and measuring the NIR power law spectral index, we are able to extend this synchrotron spectrum to higher energies to assess whether it could contribute to the X-ray flux. In XTE J1550--564, it was found \citep{russet10} that the extrapolated IR power law and X-ray power law were consistent with being the same origin, and were linearly correlated, implying that the X-rays were produced by optically thin synchrotron emission in the jet. However at higher luminosities in the hard state ($> 2 \times 10^{-3} L_{\rm Edd}$) the IR power law underpredicted the X-ray flux, and thermal Comptonization from the inflow likely produced the hard power law. In XTE J1752--223, a late X-ray re-brightening may have had the same origin \citep{russet12} as a synchrotron flare seen at optical wavelengths \citep[see also][]{chunet13}. It was also noted that there is an X-ray softening during outburst decays in some sources, at a time when the IR jet becomes bright, and this X-ray softening may be due to jet synchrotron emission \citep[see section 4.6 in][]{kaleet13}.

\begin{figure}
\centering
\includegraphics[height=\columnwidth,angle=270]{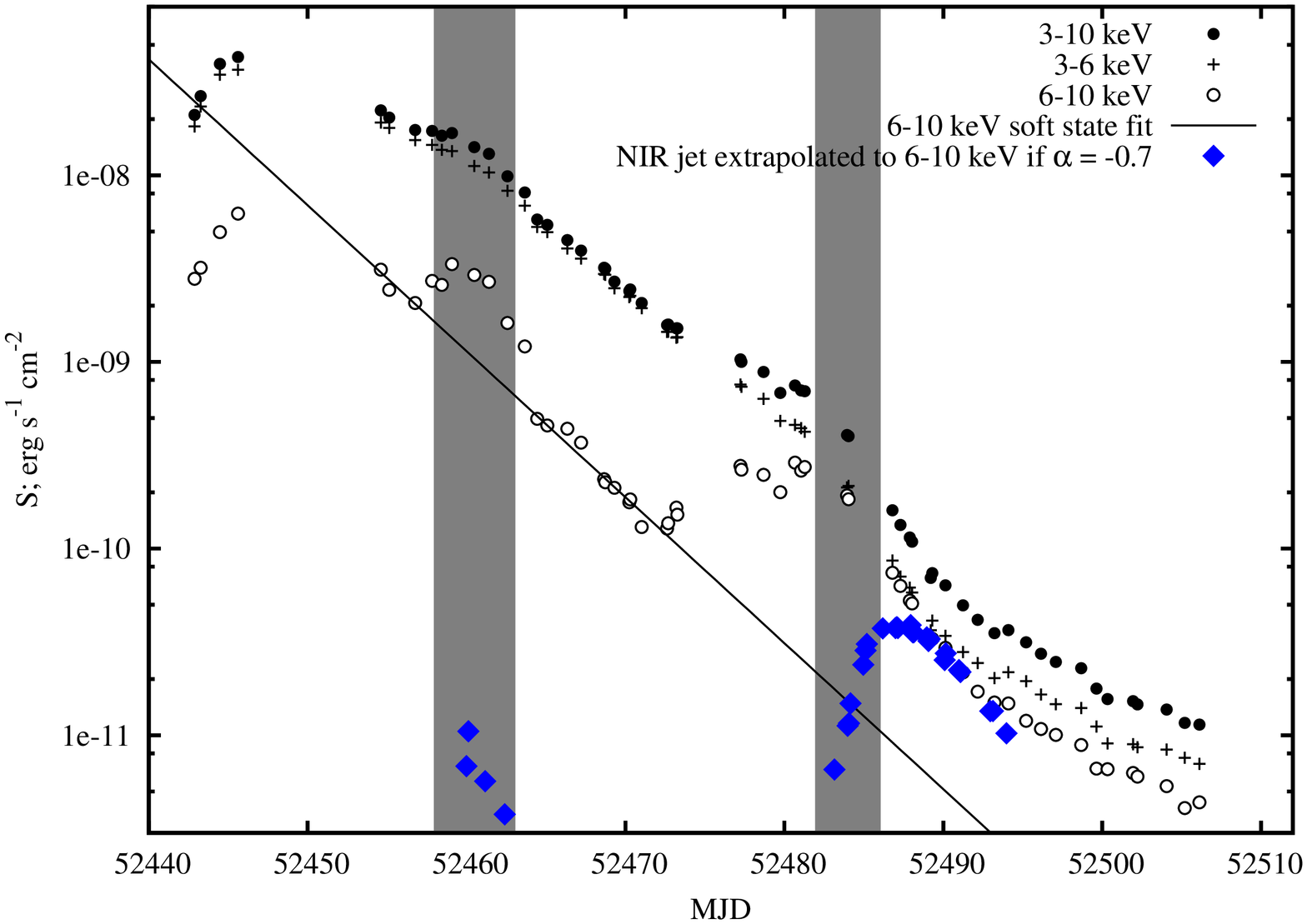}
\includegraphics[height=\columnwidth,angle=270]{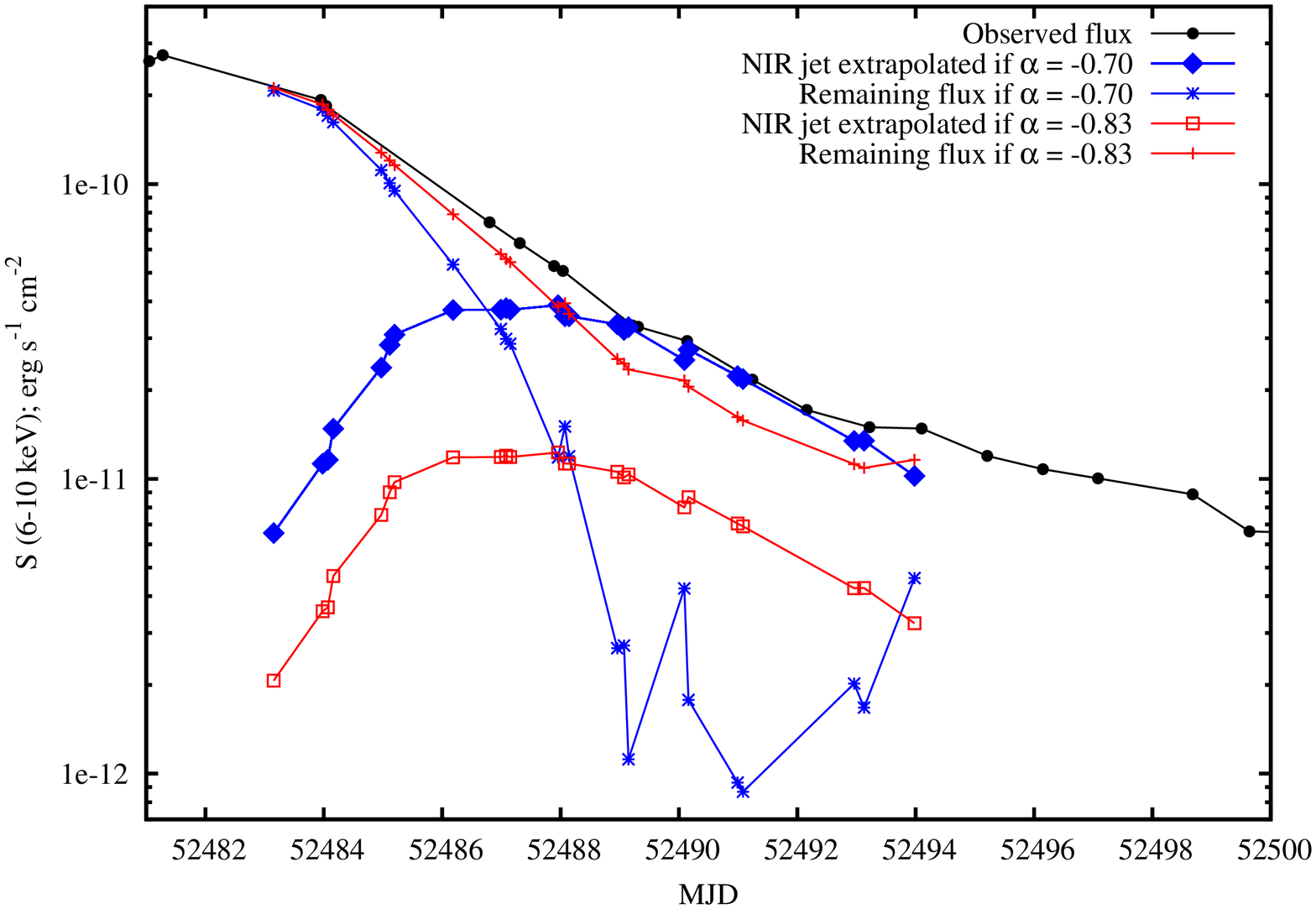}
\caption{\emph{Upper panel:} X-ray light curve in the 3--6, 6--10 and 3--10 keV energy bands. The grey shaded regions are the same as in Fig. \ref{lc1} and \ref{lcjet}. The hard X-ray increase during the soft to hard transition is clearly visible after MJD 52473. If the IR excess has an optically thin synchrotron spectrum (with $\alpha = -0.7$ from IR to X-ray energies, its light curve would be the blue filled diamonds. This IR jet synchrotron spectrum extrapolated to X-ray energies (6--10 keV) is negligible compared to the observed X-ray flux \citep[see also][]{kaleet05}, at all times except around MJD 52490, at which time it approaches the observed X-ray flux if $\alpha = -0.7$, or $\sim 30$ per cent of the X-ray flux if $\alpha = -0.83$ (see Fig. \ref{lcjetX}). \emph{Lower panel:} The 6--10 keV light curve for the jet and non-jet X-ray components, for the cases of the jet spectrum having $\alpha = -0.7$ (blue) or $\alpha = -0.83$ (red); see text for details.}
\label{lcX}
\end{figure}

In Fig. \ref{lcX} (upper panel) we plot the observed X-ray light curve (in three bands to illustrate the soft and hard components) of 4U 1543--47. The excursion to the SIMS is seen as a hard X-ray flare in the 6--10 keV light curve around MJD 52460. The transition to the hard state is visible as a hard X-ray flare starting on MJD $\sim 52473$. The transition from HIMS to hard state was completed by MJD 52483 at a luminosity of $1.42 \pm 0.19$ per cent of the Eddington luminosity \citep{vahdet19}. During the transition, the rms variability increases (Fig. \ref{lc1}) and the peak frequencies in the power spectrum are decreasing \citep{dincet12}.

Assuming the optically thin jet synchrotron spectrum extends from the IR to X-ray energies with a spectral index $\alpha = -0.7$ (an average value, see Fig. \ref{lcjet}, lower panel) or $\alpha = -0.83$ \citep[as measured from a single OIR SED on MJD 52490; see fig. 1(e) of][]{russet13a}, we can estimate the X-ray light curve of such a component. The result assuming $\alpha = -0.7$ is shown as blue filled diamonds in Fig. \ref{lcX} (upper panel). During the IR jet flare in the SIMS, the extrapolation of this IR synchrotron spectrum underpredicts the observed X-ray flux by more than two orders of magnitude (for the case of $\alpha = -0.83$ the jet flux would be even fainter). The jet cannot therefore produce the brief hard X-ray flare during the SIMS. During the outburst decay, the jet flux rises from MJD 52483 to 52487. The jet light curve underpredicts the X-ray flux and has a different light curve morphology, except for a few days around MJD 52490--52493. For these few days, it is possible that the synchrotron jet could make a significant contribution to the X-ray emission, and could dominate the X-ray flux if $\alpha = -0.7$ for the jet.

In the lower panel of Fig. \ref{lcX}, we investigate this period of the light curve further and show the jet X-ray light curve if $\alpha = -0.7$ (blue filled diamonds) or $\alpha = -0.83$ (red open squares). The observed flux (black filled circles) is also shown, and the remaining, non-jet flux is plotted for both cases (blue stars for $\alpha = -0.7$, red crosses for $\alpha = -0.83$). It is clear that if the jet has a $\alpha = -0.7$ spectrum from IR to X-ray, the jet would dominate the X-ray flux but for this to be the case, the non-jet flux (most likely the Comptonized flux from the inner flow) must fade dramatically then recover on a timescale of $\sim$ one week, in order to explain the X-ray light curve. This is unlikely, as we would expect this component to decay monotonically, and we see no reason for the Comptonized component to fade and then brighten just when the jet dominates the X-ray flux. However, if $\alpha = -0.83$, the jet flux is $\sim 3$ times fainter than the $\alpha = -0.7$ case, and the non-jet flux shows a fairly smooth decay, as expected. We can therefore conclude that it is unlikely that the jet dominates the X-ray flux, but it could produce some significant fraction ($\lesssim 30$ per cent) of the emission at this time.

\begin{figure}
\centering
\includegraphics[height=\columnwidth,angle=270]{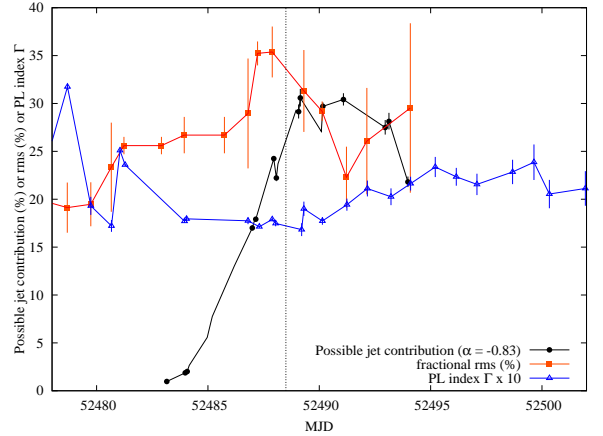}
\caption{The possible jet contribution to the X-ray flux, as a percentage (filled black circles) during outburst decay, if the jet has a power law spectrum of $\alpha = -0.83$ from IR to X-ray. The vertical line illustrates the epoch after which the jet could contribute significantly to the X-ray flux (MJD 52488.5). The X-ray fractional rms (filled orange squares) and X-ray power law index ($\times 10$; open blue triangles) are overplotted, to test for X-ray changes during this time.}
\label{lcjetX}
\end{figure}

In Fig. \ref{lcjetX} the light curve of the possible jet contribution (as a fraction of the X-ray flux) is plotted (black circles) for the case of $\alpha = -0.83$. The possible jet contribution peaks at $\sim 30$ per cent for $\sim 5$ days (for $\alpha = -0.7$ the peak would occur over the same date range but would peak at $\sim 90$--100 per cent). Overplotted is the X-ray fractional rms variability (orange squares) and the X-ray power law photon index ($\times 10$; blue triangles). At this time, the X-ray photon index was $\Gamma = 1.8$--2.1, and slightly increasing. It was $\Gamma = 1.8$ on MJD 52489, when the jet could dominate. This is close to the assumed IR to X-ray power law index, $\Gamma = 1.7$ (if $\alpha = -0.7$) so at this time the X-ray power law was consistent with being the extrapolation of the IR to X-ray power law. However, as stated above, this would require the non-jet X-ray power law to fade and recover at the same time, which is improbable. Over the next few days the power law index increased to $\Gamma = 2.1$ ($\alpha = -1.1$). This could be due to the jet component being steeper ($\alpha = -0.83$; $\Gamma = 1.83$ or the high energy turnover / cooling break in the jet spectrum may be in, or close to, the X-ray band), or the softening may be partly caused by a Galactic ridge contribution in this decay stage, especially at late times when it is faintest \citep[the ridge spectrum is estimated in][]{dunnet10}.

The rms variability shows a slight drop from just above 30 per cent to $\sim 20$ per cent around MJD 52488--52491 then increases slightly (although errors are large). It is therefore possible that the rms drop might be associated with synchrotron emission from the jet contributing to the X-ray emission, although the rms drop is gradual compared to the rapid rise of the possible jet contribution to the X-ray power law. There could be a delay due to jet spectral evolution; the optically thin--thick spectral break will be increasing in frequency during the transition, and it has been found that the IR power spectrum of GX 339--4 evolves slowly during the outburst decay, while the X-ray power spectrum evolves faster \citep{vincet19}. Alternatively, the slight drop in rms at this time could be a coincidence. At OIR wavelengths, the rms variability of the optically thin jet power law is expected to be $\sim 10$--20 per cent \citep[e.g.][]{gandet10,cadoet11,baglet18} so this is consistent with the decrease from $\sim 30$ per cent to $\sim 20$ per cent. More accurate rms measurements, and the evolution of well fitted power spectra, would be needed to confirm any jet association with an rms drop or rise, for future outbursts.

In XTE J1550--564, an increase in the X-ray photon index was seen when the jet came to dominate the X-rays, similar to here, but no X-ray timing (rms) information was available for that source \citep{russet10}. In XTE J1752--223, when the jet may have produced a late X-ray flare, there was no significant change in the photon index ($\Gamma = 1.5$) or the rms ($\sim 26$ per cent) before and during the flare \citep{russet12}. Here, in 4U 1543--47 there is a slight decrease in the rms and a smooth increase in the photon index. To summarize, no solid conclusions can be made about the jet contribution to the X-ray flux, although it is likely to be $\lesssim 30$ per cent at all times but may make a contribution up to that value around MJD $\sim 52490$. If the cooling break lies at lower frequencies in the UV, the jet will make a negligible contribution to the X-ray flux.

\section*{Acknowledgements}

We thank Robert Dunn for providing the RXTE data analysis from \citep{dunnet10} and for discussions.
We also thank Rob Fender and Tomaso Belloni for feedback on an early version of this result, and the reviewer for a constructive report that helped to improve this work. These results are based on data taken with the YALO telescope (currently operated by the SMARTS consortium), and NASA's RXTE X-ray mission.











\bsp	
\label{lastpage}
\end{document}